# The Structural Dynamics of the World Trade Center Catastrophe

**Ansgar Schneider**

Dr. rer. nat.

Bonn, Germany

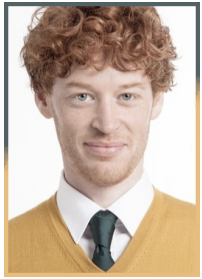

## 1. Abstract

Bažant et al. have proposed a model of a gravity-driven collapse of a tall building which collapses in a progressive-floor collapse after the failure of a single storey. The model allows the re-computation of the structural resistance of the building once the downward movement of the building has been quantified. We give a physically more sound version of the collapse model, and determine the downward movement of the North Tower of the World Trade Center. Thereby we reproduce a value for the upward resisting force during the collapse that is similar to what has been achieved by Bažant et al. for the first three seconds of the collapse. However, our method of measurement also includes data up to 9 seconds after collapse initiation. These data show a much bigger upward resistance force between 4 and 7 seconds after collapse initiation.

## 2. Introduction

### 2.1. 11th of September 2001

On the 11th of September 2001 the three skyscrapers of the World Trade Center (WTC) in New York City (NYC) were destroyed. Two of them, the Twin Towers (the North and the South Tower), had been struck by an aircraft and subsequently were aflame. Here we shall restrict our discussion to the collapse of the North Tower.

The National Institute of Standards and Technology (NIST) released a report in 2005 dealing with a hypothetical scenario about the *collapse initiation* [1]. However, an explanation of the *collapse itself* was not given by NIST, so further investigation is needed.

### 2.2. Progressive Floor Collapse

In a series of papers Bažant et al. [2,3,4] have proposed the model of a progressive floor collapse: After the failure of one storey the falling top segment of the building impacts the structure below, then floor by floor the building structure is destroyed (cp. Figure 1). The collapsing building is characterised by three segments: 1. the undestroyed top segment which keeps its height $z_0$ until the crushing front hits the ground, 2. the compacted/destroyed middle segment which is getting larger during the collapse, and 3. the still undamaged bottom segment which shrinks in height as the crushing front progresses.

The coordinate system to which we will always refer has its origin at the undestroyed tower top (cp. Figure 1). The crushing front is at position $z(t)$ at time $t$ and progresses down the building, so at collapse initiation ($t = 0$), we have $z(0) = z_0 =$ **46 m** (see Sections 2.4 and 3.1 of the supplementary material for this article[1] for a further discussion about the numerical value of $z_0$). The position of the roof at time $t$ is $x(t)$, and we assume that at each time $t$ the downward velocity is spatially constant within the two moving segments and is therefore given by $\dot{x}(t)$, the time derivative of $t \mapsto x(t)$. In other words, the top segment and the compacted segment move as a single body, which is increasing its mass and extension as the collapse progresses. The amount of compaction of the middle segment is described by the so-called compaction parameter $\lambda$, i.e. $\lambda(z(t) - z_0)$ is the height of the middle segment. We assume $\lambda$ is constant in time, i.e. each storey is compacted by the same amount. As in [5] we shall assume a numerical value of $\lambda = 0.15$.

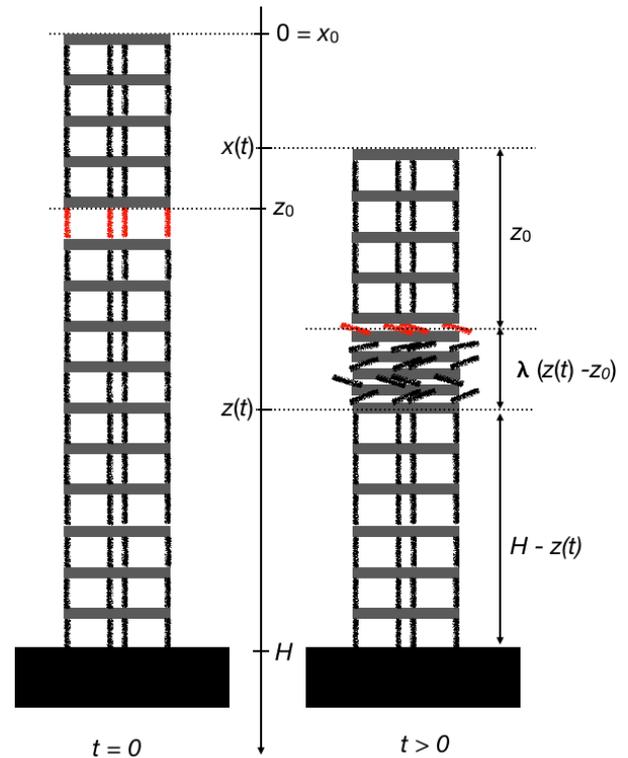

*Figure 1. Schematic illustration of a progressive floor collapse*

There is an obvious one-to-one correspondence between $x(t)$ and $z(t)$, given by $x(t) = (1-\lambda)(z(t) - z_0)$, and so the the downward velocity of the crushing front $\dot{z}(t)$ satisfies

$$\dot{x}(t) = (1-\lambda)\dot{z}(t). \tag{1}$$

With the use of the geometric relation (1) we shall now formulate momentum conservation in our setting.

---

[1] The supplementary material for this article is available online on the e-print server of the Cornell University Library [5].

## 2.3. The Crush-Down Equation

Let $\mu(\,\cdot\,)$ be the mass height-density of the undamaged building, i.e. the total mass (above the ground) of the building of height $H = 417\,\mathrm{m}$ is given by

$$\int_0^H \mu(y)\,dy, \qquad (2)$$

then the collective mass $m(z)$ of the mass of the top segment and the mass of the middle segment is given by

$$m(z) = \int_0^{z_0} \mu(y)\,dy + (1-\kappa)\int_{z_0}^z \mu(y)\,dy, \quad (3)$$

where $\kappa$ is the parameter that specifies how much building material is thrown outwards at the crushing front. We shall use a numerical value of $\kappa = 0.25$, but also remark that the actual value of $\kappa$ only has a very small numerical effect on our computations (cp. Figure 5 in [5].)

Now by (1) the downward moving momentum at time $t$ is given by $m(z(t))(1-\lambda)\dot{z}(t)$. So momentum conservation yields an ordinary differential equation for the function $t \mapsto z(t)$ which we call the Crush-Down Equation (CDE):

$$\frac{d}{dt}\Big(m(z)(1-\lambda)\dot{z}\Big) = m(z)g - F(z) - (\alpha\mu(z) - \beta)\dot{z}^2, \qquad (4)$$

wherein $g$ is NYC's acceleration of gravity; $F(z)$ is the upward directed resistance force of the crushing columns (that is the quantity in which we are interested); $\alpha, \beta$ are constants.

The origin of the gravity term and of the term involving $F(z)$ is obvious. The quadratic term with $\alpha, \beta$ can be regarded as some type of friction term (cp. the discussion in Section 1.4 of [5]). The numerical values in [4,6] are $\alpha = 0.02$, $\beta = 0.5 \cdot 10^5\,\mathrm{kg/m}$. Again the precise numerical value has only a small effect on the computation (cp. Figure 5 in [5]).

**Remark:** The above version of the CDE is sightly different from the one in [4], where a non-trivial spatial velocity profile of the middle segment is discussed. However, this is not done accurately, as it it based on unphysical assumptions. A lengthy and technical discussion of this aspect is given in [6], where it is also shown that the numerical difference of the unphysical solutions of [4] and the solutions of (4) are rather small.

## 2.4. The Damage Function $\chi(\,\cdot\,)$ and the Initial Value Problem

To model the damage of the tower let $\chi(z) \in [0,1]$ be the parameter which specifies how much the columns are weakened at $z$. I.e. $\chi(z) = 1$ means full support, and the upward force is the product $F(z) = \chi(z)F_0(z)$, where $F_0(\,\cdot\,)$ describes the resistance during the collapse without any external damage (s. below). For our numerical analysis we will consider four damaged storeys: $\chi(z) = 0.5$ if $z$ is in the range of the first failing storey (between $z_0$ and $z_0 + h$), and $\chi(z) = 0.9$ if $z$ is in the range of the three stories below (between $z_0 + h$ and $z_0 + 4h$).

Similar values have been used in [3,4].

The initial conditions for our investigation of the CDE are $z(0) = z_0$ and $\dot{z}(0) = 0$, i.e. no downward velocity at time $t = 0$. Nonetheless the propagation of the collapse is triggered by the reduction of the upward force by the damage function $\chi(\,\cdot\,)$.

## 2.5. The Shape of $\mu(\,\cdot\,)$ and $F_0(\,\cdot\,)$

The shapes of the mass density $\mu(\,\cdot\,)$ and the resistance force $F_0(\,\cdot\,)$ are specified essentially as piece-wise linear functions in Figure 2(a) of [4]. Figure 2 displays the normalised versions $\overline{\mu}(\,\cdot\,), \overline{F}(\,\cdot\,)$ of the actual quantities, i.e.

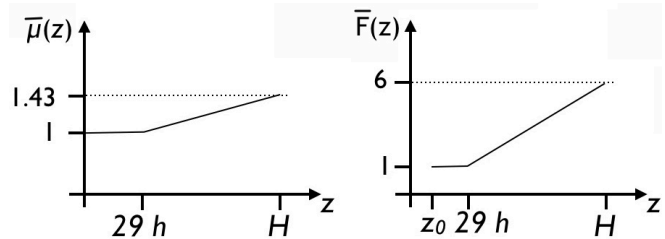

*Figure 2. Normalised average mass density and resistance force,*

h = 3.8 m = height of one storey

$\mu(z) = \mu_0 \overline{\mu}(z)$, where $\mu_0 = 0.6 \cdot 10^6$ **kg/m** is chosen such that the total mass of the tower including its 21 storeys below the ground is about 300.000 tons (cp. p. 8 of [5]).

$\overline{F}(\cdot)$ is chosen proportional to the cross-section area of the columns. So the average force is $F_0(z) = F_{av}\overline{F}(z)$, where $F_{av}$ is the average resistance force in the storeys above the 80th floor (i.e. $z < 29h$, the building had 110 storeys of height $h =$ **3.8 m**).

## 3. Empirical Data and Numerical Solution of the CDE

### 3.1. Gaining Basic Data

Plenty of video footage is available from the collapse of the tower. During the first 3.2 seconds of the collapse the roofline of the building is visible and its decent can directly be measured. Until 4.6 seconds parts of the 110 meter tall antenna on top of the building are visible and can also be used to determine the position of the roof. After 4.6 seconds the view onto the upper parts of the building is obstructed by the dust cloud. However, the collapse still can be tracked by following the well-defined and accurately downward moving crushing front or, to be more precise, the bottom part of the dust cloud, which might be ahead of the crushing front and therefore gives a lower bound for the crushing front.

For the sake of completeness the lengthy details of all of our measurements can be found in Section 2 of the supplementary material [5]. Here we shall only present the main results.

Once the measurements are done one can fit the solutions of the CDE to the empirical data. It is standard how to treat a 2nd order differential equation like the CDE and how to compute its numerical solutions. We don't comment on that any further, but for completeness the details can be found in Section 1.4 and Appendix C of [5].

### 3.2. The Roofline Measurements

At collapse initiation the elevation of the roofline is at 417 m. Then during the first 4.6 seconds of the collapse we find it at

408 m at $t = 1.64$ sec, (5)

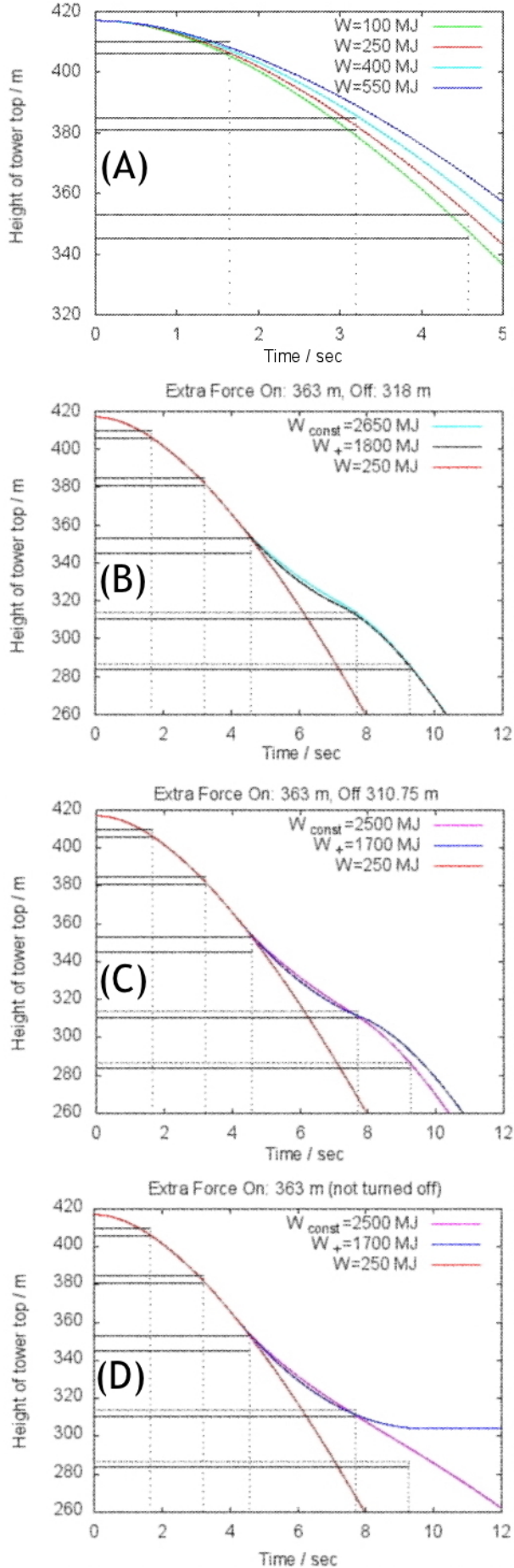

*Figure 3. Empirical results and numerical solutions of the CDE*

$$383 \text{ m at } t = 3.20 \text{ sec,} \tag{6}$$

$$349 \text{ m at } t = 4.57 \text{ sec.} \tag{7}$$

In Figure 3 (A) the black horizontal lines are the error bars of these measurements. The measured values in the middle of each pair of lines are not drawn.

The coloured graphs show the elevation of the roofline computed by the CDE, i.e. it displays the function,

$$t \mapsto H - (1 - \lambda)(z(t) - z_0), \tag{8}$$

where the solutions of the CDE are computed for an average force of

$$F_{av} = 26 \text{ MN, } 66 \text{ MN, } 105 \text{ MN, } 132 \text{ MN,} \tag{9}$$

for the green, red, cyan and blue graph, respectively. (The indicated energy values are given by $W = F_{av}h$.)

Apparently the average resistance force during the first 4.6 seconds of the collapse was on a scale of $F_{av} = 66$ MN ($= 250$ MJ/$h$).

**Remark:** This value has the same order of magnitude as the result of [4] which is $F_{av} = 100$ MN. A detailed error analysis of the difference of these two values seems to be pointless, because it is not specified which video material has been used in [4]. However, one reason for this difference is that the total mass of the tower is assumed to big in [2,3,4]. (See [5] for a discussion.)

### 3.3. Tracking the Crushing Front

We made two measurements of the crushing front and find it at

$$z(t) = 248 \text{ m at } t = 7.71 \text{ sec,} \tag{10}$$

$$z(t) = 216 \text{ m at } t = 9.25 \text{ sec.} \tag{11}$$

Then by equation (8) we can recompute the elevation of the roofline, which gives

$$312 \text{ m at } t = 7.71 \text{ sec,} \tag{12}$$

$$285 \text{ m at } t = 9.25 \text{ sec.} \tag{13}$$

Similar to (A), in Figure 3 (B), (C) and (D) the error bars of the measured values (5), (6), (7), (12) and (13) are displayed by the horizontal black lines. The upper error bars of (12) and (13) are dashed only, because the crushing front might be above the dust cloud.

In all three diagrams the red graph is the solution of the CDE for $F_{av} = 66$ **MN** as in (A). At $t = 7.71$ **sec** it misses the measured value by 40 m.

The other coloured graphs are solutions where — when the roofline is located in the intervals specified above the diagrams — an additional force is added to compute the solution, namely either

$$F_+(z) = W_+/h \, \overline{F}(z) \text{ or} \tag{14}$$

$$F_{const}(z) = W_{const}/h. \tag{15}$$

The solutions with the constant extra force are displayed only for reasons of comparison. The more informative solutions are those computed for the extra force $F_+(z)$, because this force is directly proportional to $F(z)$, i.e. it reflects the structure of the building and the sum $F_{av} + W_+/h$ can be directly compared to $F_{av}$.

The extra forces are turned on when the roof has an elevation of 363 m, but they are applied over a different length: The shortest interval is used in (B), a slightly longer interval in (C) and a half open interval in (D). The interval in (C) covers the full time between 4.57 and 7.71 seconds, i.e. the extra force therein

$$W_+/h = 1700 \text{ MJ}/3.8 \text{ m} = 450 \text{ MN} \tag{16}$$

is the minimal additional force that is necessary to reach the position at 7.71 seconds. Note that this force is sufficiently big to arrest the collapse as shown in (D).

In (B) the interval is a little shorter, but the extra forces are a little bigger. The solutions therein fit better to the empirical value at 9.25 seconds.

### 4. Discussion and Conclusion

We have coupled the model of a progressive floor collapse to the destruction of the North Tower of the WTC. The implications of this model are:

1. The possible average resistance force of the collapsing building structure was on a scale above 500 MN.

2. The de facto resistance force during the first 4.6 seconds of the collapse and after 7.7 seconds was nearly one order of magnitude lower (66 MN).

In view of diagram (D) it is of particular importance to understand which mechanism reduced the resistance of the building structure. It seems to be crucial to point out that the sometimes expressed belief that the building structure was a priori too weak to arrest the collapse after it had begun is false (even after the top segment of the building had gained a significant amount of momentum).

A thorough investigation of the collapse has to be made to clarify this important issue.